\documentclass[useAMS,usenatbib,usegraphicx]{mn2e}
\usepackage{amsmath}
\usepackage{amsfonts}
\usepackage{amssymb}
\begin{document}
\title[BOOMERanG Constraints on Primordial Non-Gaussianity]{BOOMERanG Constraints on Primordial Non-Gaussianity from Analytical Minkowski Functionals}

\author[P.\ Natoli {\it et al.}]
{P. Natoli$^{1,2}$\thanks{\tt paolo.natoli@gmail.com}, 
G. De Troia$^{1}$, 
C. Hikage$^{3,4}$, 
E. Komatsu$^{5}$, 
M. Migliaccio$^{1}$, \newauthor
P. A. R. Ade$^{4}$, 
J. J. Bock$^{6}$, 
J. R. Bond$^{7}$, 
J. Borrill$^{8,9}$, 
A. Boscaleri$^{10}$,\newauthor
C. R. Contaldi$^{11}$, 
B. P. Crill$^{6}$, 
P. de Bernardis$^{12}$, 
G. de Gasperis$^{1}$,\newauthor
A. de Oliveira-Costa$^{13}$, 
G. Di Stefano$^{14}$, 
E. Hivon$^{15}$, 
T. S. Kisner$^{8,9}$,\newauthor
W. C. Jones$^{16}$, 
A. E. Lange$^{17}$, 
S. Masi$^{12}$, 
P. D. Mauskopf$^{4}$, 
C. J. MacTavish$^{18}$,\newauthor
A. Melchiorri$^{12,19}$, 
T. E. Montroy$^{20}$, 
C. B. Netterfield$^{21}$, 
E. Pascale$^{21}$, \newauthor
F. Piacentini$^{12}$, 
G. Polenta$^{12,22,23}$, 
S. Ricciardi$^{8,9}$, 
G. Romeo$^{14}$, 
J. E. Ruhl$^{20}$,\newauthor
M. Tegmark$^{13}$, 
M. Veneziani$^{12}$ 
and N. Vittorio$^{1}$ \\
\\
$^{1}$ Dipartimento di Fisica, Universit\`a di Roma ``Tor Vergata'',
Via della Ricerca Scientifica, 1 I-00133 Roma, Italy\\
$^{2}$ INFN, Sezione di Tor Vergata, Roma, Italy\\
$^{3}$ Department of Astrophysical Sciences, Princeton University, Peyton Hall, Princeton NJ 08544, USA \\
$^{4}$ School of Physics and Astronomy, Cardiff University, Cardiff, CF24 3AA \\
$^{5}$ Texas Cosmology Center, University of Texas at Austin, 1 University Station, C1400, Austin, TX 78712, USA\\
$^{6}$ Jet Propulsion Laboratory, Pasadena, CA, USA\\
$^{7}$ Canadian Institute for Theoretical Astrophysics, University of Toronto, Toronto, Ontario, Canada\\
$^{8}$ Computational Research Division, Lawrence Berkeley National Laboratory, Berkeley, CA, USA\\
$^{9}$ Space Sciences Laboratory, UC Berkeley, CA, USA\\
$^{10}$ IFAC-CNR, Firenze, Italy\\
$^{11}$ Theoretical Physics Group, Imperial College, London\\
$^{12}$ Dipartimento di Fisica, Universit\`a La Sapienza, Roma, Italy\\
$^{13}$ Department of Physics, MIT, Cambridge, MA 02139, USA\\
$^{14}$ Istituto Nazionale di Geofisica e Vulcanologia, 00143 Rome, Italy\\
$^{15}$ Institut d'Astrophysique, Paris, France\\
$^{16}$ Department of Physics, Princeton University, Princeton, NJ 0854, USA\\
$^{17}$ Observational Cosmology, California Institute of Technology, Pasadena, CA, USA\\
$^{18}$ Astrophysics Group, Imperial College, London\\
$^{19}$ INFN, Sezione di Roma 1, Roma, Italy\\
$^{20}$ Physics Department, Case Western Reserve University, Cleveland, OH, USA\\
$^{21}$ Physics Department, University of Toronto, Toronto, Ontario, Canada\\
$^{22}$ ASI Science Data Center, c/o ESRIN, 00044 Frascati, Italy\\
$^{23}$ INAF - Osservatorio Astronomico di Roma, Monte Porzio Catone, Italy\\
}




\maketitle
\begin{abstract}
We use Minkowski Functionals (MF) to constrain a primordial
non-Gaussian contribution to the CMB intensity field as observed in
the 150 GHz and 145 GHz BOOMERanG maps from the 1998 and 2003 flights,
respectively, performing for the first time a joint analysis of the
two datasets.  A perturbative expansion of the MF formulae in the
limit of a weakly non-Gaussian field yields analytical formulae,
derived by \citet{hikage}, which can be used to constrain the coupling
parameter $f_{\rm NL}$ without the need for non-Gaussian
simulations. We find $-1020<f_{\rm NL}<390$ at $95\%$ CL,
significantly improving the previous constraints by \citet{detroia} on
the BOOMERanG 2003 dataset.  These are the best $f_{\rm NL}$ limits to
date for suborbital probes.
\end{abstract}
\section{Introduction}

Detection of non-Gaussian signals in the Cosmic Microwave Background
anisotropy pattern can be of significant help in discriminating
between different inflationary models.  The most simple inflationary
models based on single-rolling scalar fields predict very small
deviations from Gaussianity that cannot be usefully constrained by
present day experimental efforts \citep{bartolo}.  However,
multi-field inflationary models and other alternative scenarios allow
for more relevant non-Gaussian contribution that could be in principle
detected by current and forthcoming missions
\citep{bernardeau,lyth}. In this paper we consider only the so-called
local form for primordial non-Gaussianity, which can be parametrized
by means of a quadratic term in Bardeen's curvature perturbations
$\Phi$
\citep{1990PhRvD..42.3936S,1994ApJ...430..447G,2000MNRAS.313..141V,2001PhRvD..63f3002K}:
\begin{eqnarray}\label{eqn:fnl}
\Phi(\mathbf{x})=\Phi_G(\mathbf{x})+f_{\rm NL} \left[\Phi_G(\mathbf{x})^2 -
\left\langle \Phi_G(\mathbf{x})^2 \right\rangle \right],
\end{eqnarray}
where $\Phi_G$ is a zero mean, Gaussian random field and $f_{\rm NL}$
is the coupling parameter that characterizes the amplitude of
primordial non-Gaussianity. At present, the most stringent limits on
$f_{\rm NL}$ are derived from the WMAP five year analysis at
$-4<f_{\rm NL}<80$ ($95\%$ CL) using an optimal (i.e.\
minimum-variance) bispectrum based estimator \citep{smith}. Many
teams have further analysed the WMAP dataset to yield constraints on
$f_{\rm NL}$ using a plethora of tests, including wavelet based
estimators: see e.g.\ \citet{curto09,rudjord,2009MNRAS.tmp..682P} and references
therein. All $f_{\rm NL}$ limits to date are compatible with a
Gaussian hypothesis. \citet{yadav} claimed a measure of a positive
$f_{\rm NL}$ at above $99.5\%$ CL in the WMAP three-year data using a
bispectrum based statistics; however, their claimed signal has not
been confirmed by the WMAP five-year analysis \citep{komatsu,smith}.

On the other hand, several groups have also investigated specific
signatures in the WMAP data, typically induced by low resolution
features such as anomalous spots, reporting high significance yet
unmodelled detection of non-Gaussianity
\citep{creminelli,cruz,vielva,eriksen,park,2009PhRvL.102m1301R}.

CMB suborbital experiments have also delivered $f_{\rm NL}$
constraints, particularly MAXIMA \citep{santos}, VSA \citep{smith04},
BOOMERanG \citep{detroia} and ARCHEOPS \citep{curto}.  Although such
limits are weaker than those based on WMAP, they probe a range of
angular scale that will not be accessible to space borne observation
until the advent of
\textsc{Planck}~\footnote{http://www.rssd.esa.int/index.php?project=planck}.
Among suborbital probes, \citet{detroia} set the most stringent
$f_{\rm NL}$ constraints to date at $-800<f_{\rm NL}<1050$ ($95\%$ CL)
from BOOMERanG 2003 (hereafter B03) dataset using a pixel space
statistics based on Minkowski functionals. Such constraints were
obtained using a reference Monte Carlo set composed of non-Gaussian
CMB maps.

In this paper we revisit the $f_{\rm NL}$ analysis of the BOOMERanG
dataset. We employ a larger dataset that also includes the BOOMERanG
1998 (hereafter B98) data, allowing for a larger sky coverage and
improved signal to noise. Furthermore, we apply a different, harmonic
based, Minkowski functional code that overcomes a weakness of the
previous B03 analysis, which used a flat sky approximation to compute
the functionals. Finally, we employ the perturbative approach
developed by \citet{hikage} to quantify the contribution of primordial
non-Gaussianity to MF. \citet{hikage1} successfully applied the
perturbative method to WMAP data without need of a large set of
non-Gaussian simulations.
 
The plan of this paper is as follows. In Section \ref{1} we briefly
describe the BOOMERanG experiment and the two datasets it has produced
as well as our data analysis pipeline. In Section \ref{2} we apply the
perturbative formulae to compute the MFs of the data and Gaussian
Monte Carlo simulation maps. Furthermore in Section \ref{3} we
constrain $f_{\rm NL}$ and in section \ref{4} we draw our main
conclusions.

\section{The B98 and B03 datasets}
\label{1}

BOOMERanG was launched for the first time from Antarctica in December
1998. It has observed the sky for about 10 days, centering a target
region at RA $\sim 5h$ and DEC $\sim -45^{\circ}$ that is free of
contamination by thermal emission from interstellar dust. BOOMERanG
mapped this region scanning the telescope through $60^{\circ}$ at
fixed elevation and at constant speed. At intervals of a few hours the
telescope elevation was changed in order to increase the sky coverage
\citep{crill}. The survey region aimed at CMB observations is $\sim
3\%$ of the sky or $\sim 1200$ square degrees \citep{ruhl}.  The data
were obtained using 16 spider-web bolometric detectors sensitive to
four frequency bands centered at 90, 150, 240 and 410 GHz. Here we restrict
ourselves to the 150 GHz data that have the most advantageous
combination of sensitivity and angular resolution to target the CMB
fluctuations.  The analysis of B98 dataset produced the first high
signal to noise CMB maps at sub-degree resolution and one of the first
high confidence measurements of the first acoustic peak in the CMB
anisotropy angular spectrum \citep{debe}. The Gaussianity of this dataset 
has been constrained in both pixel and harmonic space \citep{polenta,detroia03}.

The B03 experiment has been flown from Antarctica in 2003. Contrarily
to B98, B03 was capable of measuring linear polarization other than
total intensity \citep{piacentini,montroy,jones,mactavish}.  It has
observed the microwave sky for $\sim 7$ days in three frequency bands,
centered at 145, 245 and 345 GHz. Here we use only the 145 GHz data,
for reasons analogous to B98.  These have been gathered with
polarization sensitive bolometers (PSB), i.e.\ bolometers sensitive to
total intensity and a combination of the two Stokes linear
polarization parameters $Q$ and $U$ \citep{jones_psb}. The analysis of
the dataset has produced high quality maps \citep{masi} of the southern
sky that have been conveniently divided in three regions: a ``deep''
(in terms of integration time per pixel) survey region ($\sim 90$
square degrees) and a ``shallow'' survey region ($\sim 750$ square
degrees), both at high Galactic latitudes, as well as a region of
$\sim 300$ square degrees across the Galactic plane. The deep region
is completely embedded in the shallow region.

In this paper we apply a pixel mask to select a larger effective sky
region than the one used in \citet{detroia}. We have been extremely
careful in choosing this sky cut, rejecting regions potentially
contaminated by foreground emission, which shows up clearly in the B98
higher frequency maps, and pixels falling too close to the edge of the
survey region, which exhibit low signal to noise and potential visual
artefacts. The final cut we use covers about $980$ square degrees or
$2.4\%$ of the sky. This should be compared with the $\sim 700$
($1.7\%$ of the sky) employed for \citet{detroia}, which only used
B03, and with the $1.2 \%$ and $1.8 \%$ of the sky selected
respectively for the B98 analyses of \citet{polenta} and \citet{detroia03}.
The sky mask employed in this paper is the largest ever used for BOOMERanG
non-Gaussianity studies.

\begin{figure*}
\begin{center}
\includegraphics[angle=90,width=12cm]{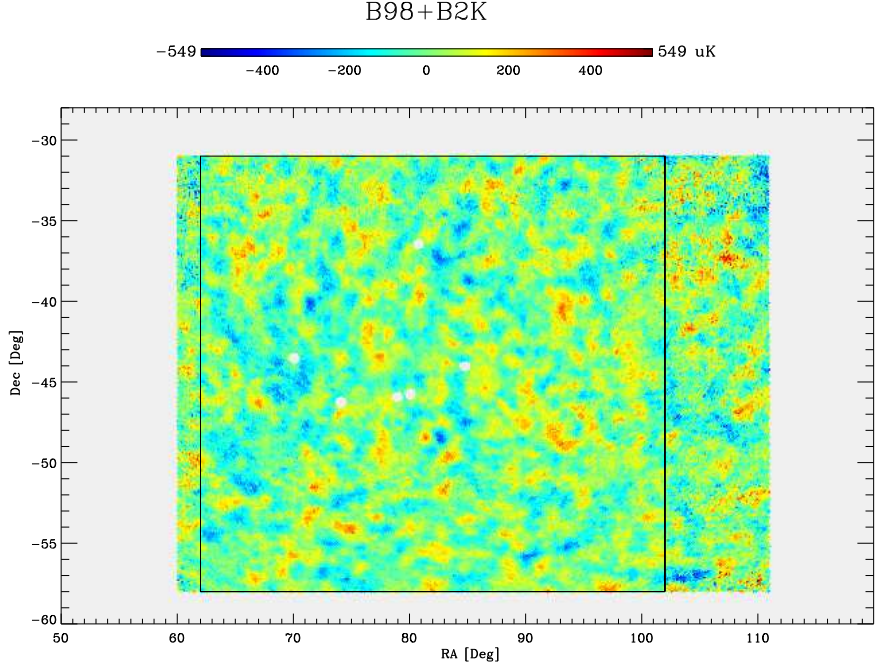}
\caption{The CMB field as seen in the B2K + B98 map, in the sky cut
used for the analysis presented here. The solid line shows the
boundary of the region taken in consideration for the analysis in
\citet{detroia}.}
\label{b98+b03_map}
\end{center}
\end{figure*}

We analysed the temperature ($T$) data map reduced jointly from eight
PSB at 145 GHz~\citep{masi} for the B03 dataset and the $T$ map
obtained from the best three of the six 150 GHz channels for
B98. While we do not consider here the Stokes $Q$ and $U$ polarization
maps, the B03 temperature map has been marginalized with respect to
linear polarization. The maps have been produced with ROMA, an
iterative generalized least square (GLS) map making code specifically
tuned for BOOMERanG analysis \citep{natoli,degasperis}. To work, the
GLS map maker needs to know the detectors' noise power spectral
densities, which is estimated directly from flight data using an
iterative procedure. In the case of B03, where cross-talks among PSB
located in the same feed horn are significant, we have also estimated
the corresponding noise cross-spectra \citep{masi}.  The timelines have
been carefully flagged to exclude unwanted data; for B98, only the
more conservative part of the scan surveyed at 1 degree per second
(d.p.s.) is used \citep{crill} while for both datasets we have flagged
all of the turn-around data.  Once the B98 and B03 maps are produced,
we obtain a single data map by noise weighting the two. In doing so we
treat the residual noise left in the map as white. This choice is
motivated by a property of the GLS map making procedure, which is very
effective in suppressing the level of noise correlations in the
data. The noise level in the B98 map roughly equals that in the B03
shallow region: at $6.7^\prime$ the noise r.m.s.\ is about 40 $\mu$K
per pixel. (While the B98 flight actually lasted longer than B03, we
consider only three channels and the 1 d.p.s.\ part of the scan here.)
The noise r.m.s.\ in the B03 deep region is $\sim 10\mu$K for
$6.7^\prime$ pixels. The joint B03/B98 map we obtain is shown in
Fig.~\ref{b98+b03_map}, in the sky cut employed for the analysis
hereafter.

To probe CMB non-Gaussianity it is important to keep under control
contaminations from astrophysical foregrounds, whose pattern is
markedly non-Gaussian. In the region selected here, foreground
intensity is known to be negligible with respect to the cosmological
signal \citep{masi}. We have masked all detectable sources in the
observed field.
To assess the robustness of our tests of Gaussianity we used a set of
1000 Monte Carlo simulation maps that mimic both the B03 and the B98 data. To
produce these simulations, the following scheme is employed. The
Gaussian CMB sky signal is simulated using the cosmological parameters
estimated from the WMAP 1-year data \citep{hinshaw} which fits well the
BOOMERanG temperature power spectrum. This signal is smoothed
according to the measured optical beam and synthesized into a
pixelized sky map, using HEALPix routines \citep{healpix}. Taking into
account the B03 and B98 scanning strategy, the signal map is projected
onto eight timestreams, one for each 145 GHz detector, for B03 and
onto three timestreams for the B98 150 GHz channels we consider
here. Noise only timestreams are also produced, as Gaussian
realizations of each detector's noise power spectral density,
iteratively estimated from flight data as explained above, taking into
account cross talks between detectors in the case of B03. The signal
and noise timelines are then added together. To reduce the simulated
timelines, we follow the same steps performed when analysing real
data: the timelines are then reduced with the ROMA mapmaking code
replicating the actual flight pointing and transient flagging, to
produce T maps jointly from three B98 channels and T, Q and U maps
jointly from all eight B03 channels. We enforce that the map making
procedure is applied to simulation and observational data following the same
steps.

It is worth noticing that in this paper the B98 and B03 data have been
used to produce a joint map for the first time.

\section{Perturbative approach to Minkowski functional for a weakly non-Gaussian CMB field}
\label{2}

In the previous paper \citep{detroia} we have also applied simple pixel
based analysis (specifically, the normalized skewness and kurtosis) to
the B03 observed field. Here we restrict ourselves to three Minkowski Functionals
generally defined in two-dimensional maps: fraction of area $V_0$,
total circumference $V_1$, and Euler Characteristic $V_2$.  We measure
the MFs for CMB temperature maps above the threshold density $\nu$,
defined as the temperature fluctuation $f\equiv\Delta T/T$ normalized
by its standard deviation $\sigma\equiv \langle
f^2\rangle^{1/2}$. Following the formalism by \citet{matsubara} and
\citet{hikage}, we can write the analytical formula for the $k$-th
Minkowski functional of weakly non-Gaussian fields as
\begin{eqnarray}
\label{eq:mf_pb}
V_k(\nu)&=&\frac{1}{(2\pi)^{(k+1)/2}}\frac{\omega_2}{\omega_{2-k}\omega_k}
\left(\frac{\sigma_1}{\sqrt2\sigma_0} \right)^k e^{-\nu^2/2}\\ \nonumber
&&\left\{H_{k-1}(\nu)+\left[\frac{1}{6}S^{(0)}H_{k+2}(\nu)
+\frac{k}{3}S^{(1)}H_{k}(\nu)+\right.\right. \\ \nonumber
&&\left.\left.\frac{k(k-1)}{6}S^{(2)}H_{k-2}(\nu)\right]\sigma_0
+\mathbf{O}(\sigma_0^2) \right\},
\end{eqnarray}
where $H_n(\nu)$ are the Hermite polynomials and
$\omega_k=\pi^{k/2}\Gamma(k/2+1)$ gives $\omega_0=1$, $\omega_1=2$,
and $\omega_2=\pi$. The $S^{(i)} (i=0, 1, 2)$ are skewness parameters,
defined by
\begin{eqnarray}
S^{(0)}&\equiv& \frac{\langle f^3\rangle}{\sigma_0^4}, \\
S^{(1)}&\equiv& -\frac{3}{4}\frac{\langle f^2(\nabla^2
f)\rangle}{\sigma_0^2\sigma_1^2}, \\ S^{(2)}&\equiv& -3\frac{\langle
(\nabla f)\cdot(\nabla f)(\nabla^2 f)\rangle}{\sigma_1^4}.
\end{eqnarray}
The variances $\sigma_j^2 (j=0, 1)$ are calculated from $C_\ell$ as
\begin{eqnarray}
\label{eq:var}
\sigma_j^2 \equiv \frac{1}{4\pi}\sum_{\ell}(2\ell+1)\left[\ell(\ell+1)\right]^jC_{\ell}W_{\ell}^2,
\end{eqnarray}
where $W_{\ell}$ is a window function that includes the experiment's
effective optical transfer function (assumed circularly symmetric) and
low-$\ell$ and high-$\ell$ cut-off as well as the filter function due
to pixelization effects. Expanding the skewness parameters into
spherical harmonics and using the reduced bispectrum
$b_{\ell_1\ell_2\ell_3}$ as a function of $f_{\rm NL}$
\citep{2001PhRvD..63f3002K}, we get:
\begin{eqnarray}
S^{(0)}&=&\frac{3}{2\pi\sigma_0^4}\sum_{2\le\ell_1\le\ell_2\le\ell_3}I^2_{\ell_1\ell_2\ell_3}b_{\ell_1\ell_2\ell_3}W_{\ell_1}W_{\ell_2}W_{\ell_3}, \\
S^{(1)}&=&\frac{3}{8\pi\sigma_0^2\sigma_1^2}\sum_{2\le\ell_1\le\ell_2\le\ell_3}[\ell_1(\ell_1+1)+\ell_2(\ell_2+1)\\ \nonumber
&&+\ell_3(\ell_3+1)]
I^2_{\ell_1\ell_2\ell_3}b_{\ell_1\ell_2\ell_3}W_{\ell_1}W_{\ell_2}W_{\ell_3}, \\ \nonumber
S^{(2)}&=&\frac{3}{4\pi\sigma_1^4}\sum_{2\le\ell_1\le\ell_2\le\ell_3}\{[\ell_1(\ell_1+1)+\ell_2(\ell_2+1) \\ 
&&-\ell_3(\ell_3+1)]\ell_3(\ell_3+1)+({\rm cyc.})\}  \\ 
&&\times I^2_{\ell_1\ell_2\ell_3}b_{\ell_1\ell_2\ell_3}W_{\ell_1}W_{\ell_2}W_{\ell_3},
\end{eqnarray}
where
\begin{eqnarray}
I_{l_1l_2l_3}\equiv
\sqrt{\frac{(2l_1+1)(2l_2+1)(2l_3+1)}{4\pi}}
\left(
\begin{array}{ccc}l_1&l_2&l_3\\0&0&0\end{array}
\right).
\end{eqnarray}

In the above theoretical predictions we assume a $\Lambda$CDM model
with the cosmological parameters at the maximum likelihood peak from
WMAP 1-year data \citep{spergel}: $\Omega_{\rm b}=0.043$, $\Omega_{\rm
cdm}=0.21$, $\Omega_{\Lambda}=0.74$, $H_0=72$km s$^{-1}$ Mpc$^{-1}$,
$n_{\rm s}=0.96$, and $\tau=0.11$. The amplitude of primordial
fluctuations has been normalized to the first peak amplitude of the
temperature power spectrum, $\ell(\ell+1)C_{\ell}/(2\pi)=(74.7 {\rm
\mu K})^2$ at $\ell=220$ \citep{page}.  

We compute the MFs of the pixelized maps by integrating a combination
of first and second angular derivatives of the temperature over the
sky, as described in Appendix A.1. of \cite{hikage}.  The threshold
density $\nu$ is set in the range $-3.6$ to $3.6$, assuming 19 evenly
spaced grid points.  For our analysis we use maps at HEALPix
\citep{healpix} resolution of $\sim 13^{\prime}$
($N_\mathrm{side}=256$) and $\sim 7^{\prime}$
($N_\mathrm{nside}=512$).

\begin{figure*}
\begin{center}
\includegraphics[width=18cm]{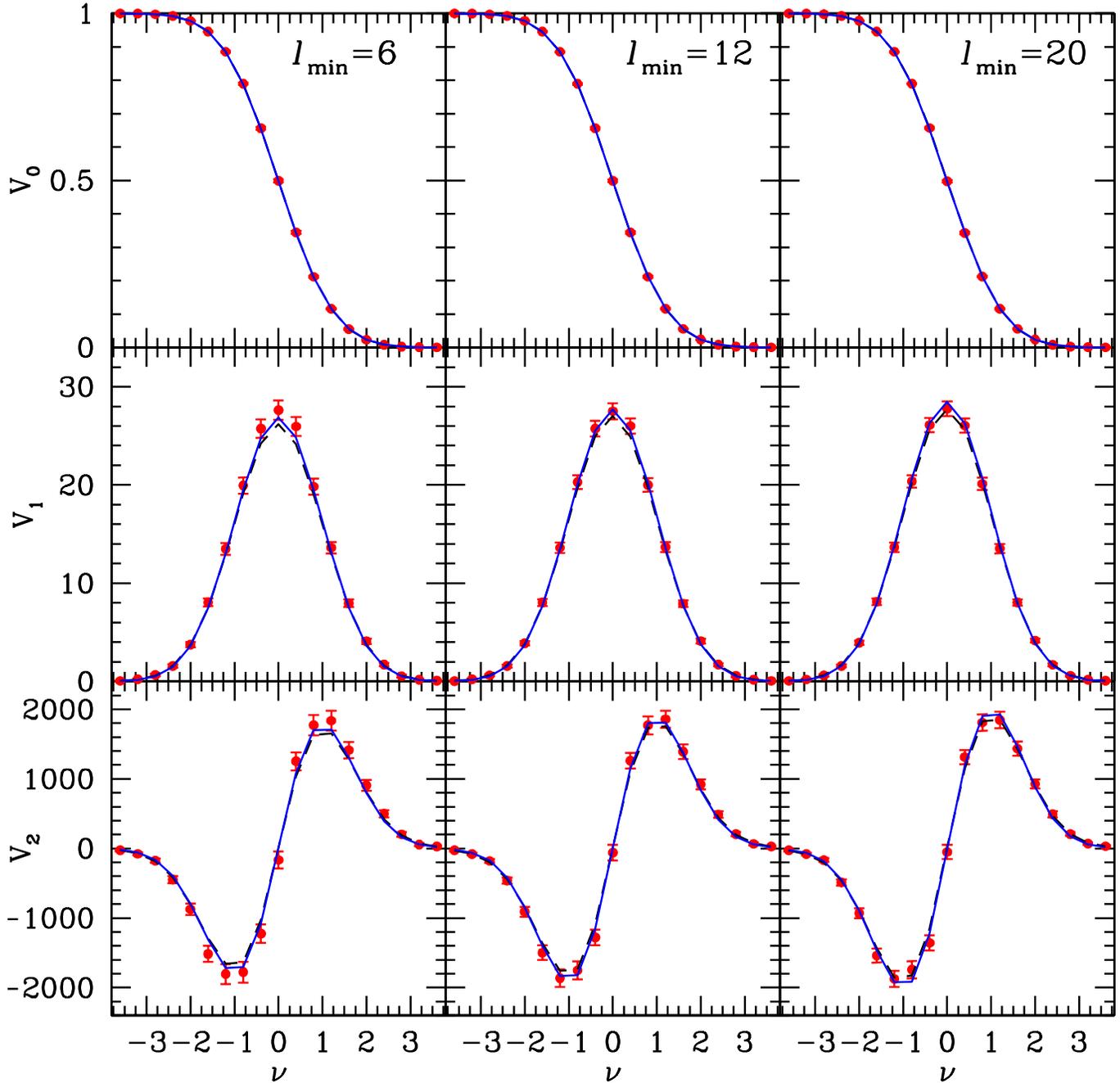}
\caption{The plots show, from top to bottom, $V_0$, $V_1$ and $V_2$
computed from the BOOMERanG data (filled circles), from analytical
formulae (solid line, here computed with the best fit $f_{\rm NL}$
value for each functional) and from Gaussian Monte Carlo simulations
(dashed lines). The columns refer to different choices of the low $\ell$
cut (see text), $\ell_{\rm min}=$ 6, 12 and 20 (from left to
right). The error-bars for the data are derived as the 1-$\sigma$
standard deviation of our Monte Carlo simulations. The pixel size of
the map is $\sim 7^{\prime}$ (HEALPix $N_\mathrm{side}=512$).}
\label{fig2}
\end{center}
\end{figure*}

\section{Constraints on primordial non-Gaussianity}
\label{3}

We define a ``joint'' estimator by grouping the $V_k$'s in a single,
$57$ elements data vector $V_J \equiv\{V_0(\nu=-3.6),
V_0(\nu=-3.2),..., V_0(\nu=3.6),V_1(\nu=-3.6),...,V_2(\nu=3.6)\}$.  We
want now to constrain the $f_{\rm NL}$ parameter and estimate its best
fit value. Starting from analytical formulae we can calculate the
non-Gaussian part of the MFs using the equation (\ref{eq:mf_pb}),
i.e.:
\begin{eqnarray}\label{eq:deltav}
\Delta V_J(f_{\rm NL})=V_J(f_{\rm NL})-V_J(f_{\rm NL}=0).
\end{eqnarray}
We can then estimate our final non-Gaussian predictions as:
\begin{eqnarray}\label{ng_predictions}
\tilde{V}_J(f_{\rm NL})=\bar{V}_J(f_{\rm NL}=0)+\Delta
V_J(f_{\rm NL}),
\end{eqnarray}
where $\bar{V}_J(f_{\rm NL}=0)$ is the average MF computed from our
Gaussian Monte Carlo simulations. The reason for this choice is that
the Monte Carlo average provides an accurate estimate of the MFs,
accounting for instrumental and coverage effect. Finally we perform
a chi-square analysis by measuring
\begin{eqnarray}\label{chi2}
\chi^2&=&\sum_{JJ^{\prime}}[V^{\rm B98+B03}_J(\nu)-\tilde{V}_J(f_{\rm NL})]
C^{-1}_{J,J^\prime}[V^{\rm B98+B03}_{J^\prime}-\tilde{V}_{J^\prime}(f_{\rm
NL})], \nonumber \\
&&
\end{eqnarray}
where $V^{\rm B98+B03}_J$ denote the MFs for the joint B98 and B03
map. This expression is used to derive constraints for $f_{\rm NL}$
and for our goodness of fit analysis. The full covariance matrix
$C_{J,J^{\prime}}$ is also estimated from Gaussian Monte Carlo
simulations. We have verified that, when computing its matrix
elements, one needs to take into account the correlations among
different functionals not to incur in biased constraints.  In Fig.
\ref{fig2} we plot each MF of the B98 and B03 data compared with the
theoretical predictions with the best fit value of $f_{\rm NL}$ for
each MF. The error-bars are derived as 1-$\sigma$ deviations estimated
from 1000 Gaussian maps with correlated noise.

We study the effect that neglecting the contribution of a range of
multipoles $\ell$ has on this analysis. A low $\ell$ cut is necessary
since we are dealing with data from a suborbital experiment, which
has not been designed to measure large angular scales. These cannot be
constrained properly, firstly because of the limited angular extension
of the region surveyed, and secondly because timeline filtering is
applied to the data to suppress contribution from low frequency noise
and systematics. The filters are applied during the map making stage
at $\sim 70$~mHz both for B03 and for B98 \citep{masi,crill}. While we
apply the same filters in our simulations, the amount of low $\ell$
power in the latter is somewhat different from those exhibited by the
data. This happens because the GLS map maker is more efficient in
recovering the large scale structure from the simulations, where we
only add Gaussian noise with random phases, than from the data where
the noise has a more complex structure. To account for this effect, we
exploit one degree of freedom allowed by the harmonic analysis to MF
here pursued: specifically, we set to zero all the power in the map
below $\ell_{\rm min}$. The left to right panels of Fig.~\ref{fig2}
refer to three different choices: $\ell_{\rm min}=$6, 12 and 20
respectively. The $\ell_{\rm min} = 6$ cut is the natural one that
would arise due to limited sky coverage but in this case the MFs from
data and simulations do not agree well for $V_1$ and $V_2$ (see Fig
\ref{fig2}). The agreement is much better for $\ell_{\rm min} = 12$
and $\ell_{\rm min} = 20$ with no appreciable difference between the
two. This comes with little surprise, since a telescope scan speed in
the range $0.5^\circ/\mathrm{s}$ to $1^\circ/\mathrm{s}$, both of
which have been employed in the dataset we consider, effectively high
pass filters the data in the range $10\lesssim\ell\lesssim 20$. (Note,
however, that timeline filtering has an anisotropic effect on the sky,
due to the nature of the scanning strategy employed for BOOMERanG.)
 
On the other hand, it is advantageous to consider also a high $\ell$
cut $\ell_{\rm max}$ to probe how the decreasing signal to noise level
can affect $f_{\rm NL}$ constraints. For the data, this can be done by
varying the HEALPix resolution parameter $N_\mathrm{side}$ which is
linked to $\ell_{\rm max}$ in the spherical harmonic transform. We
focus in the following on both $N_\mathrm{side}=256$ with $\ell_{\rm
max}=512$ and $N_\mathrm{side}=512$ with $\ell_{\rm max}=1000$. Our dataset is
signal dominated at $\ell \simeq 500$ while begins to be noise dominated
at $\ell \simeq 1000$ \citep{jones}.
 
\begin{figure*}
\begin{tabular}{c}
\resizebox{8.7cm}{!}{\includegraphics{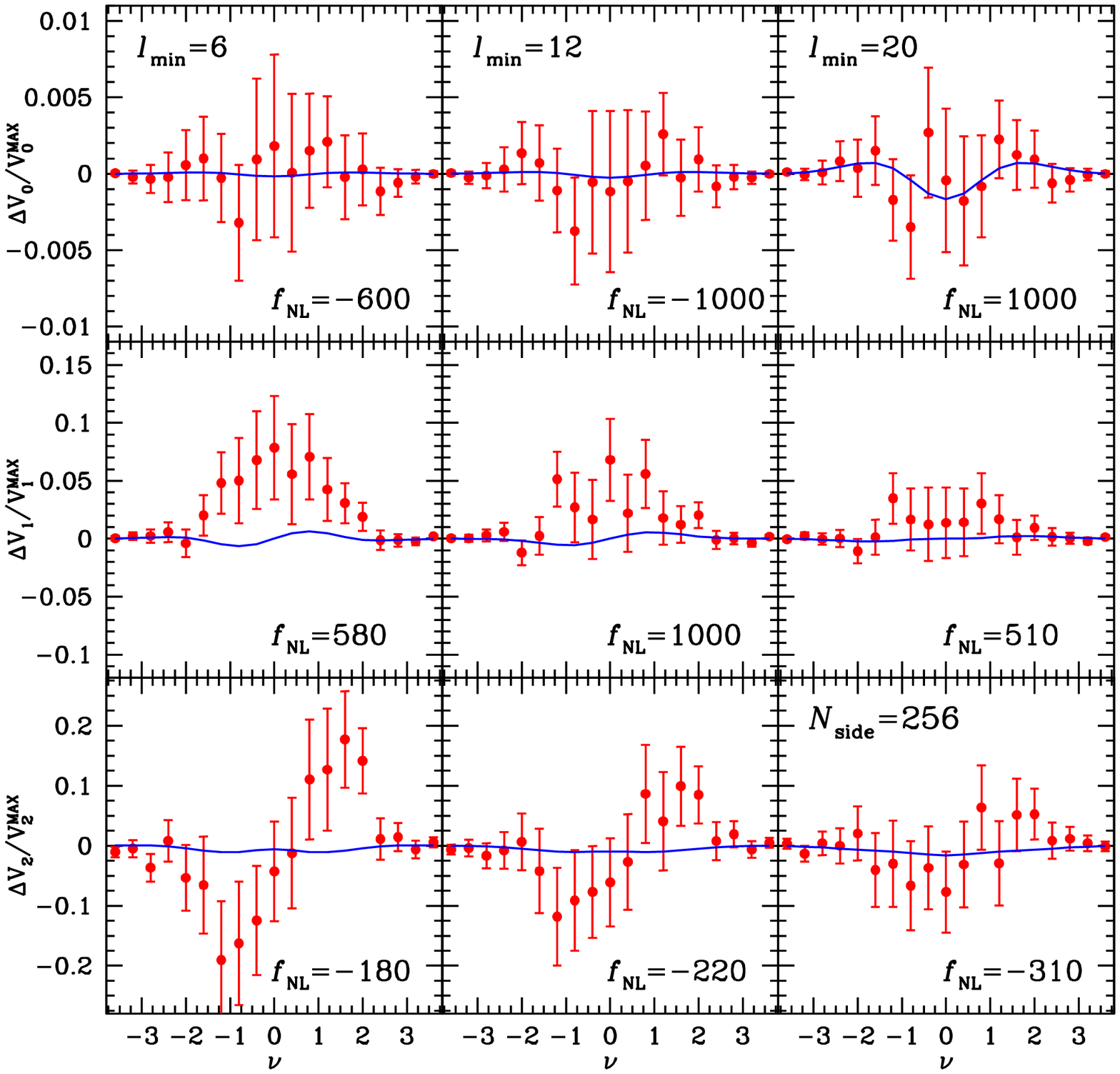}}
\hspace{0.3cm}
\resizebox{8.7cm}{!}{\includegraphics{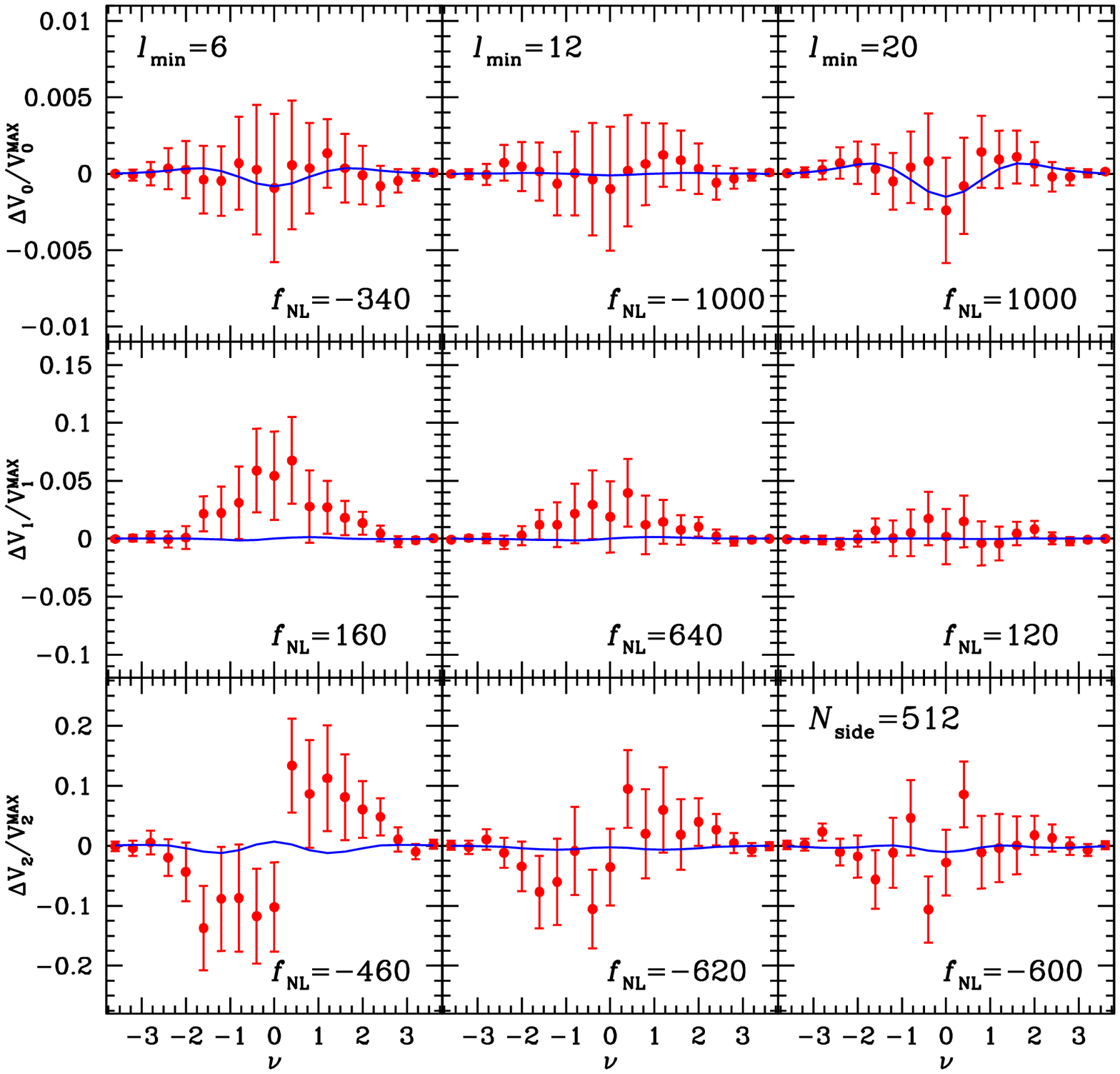}}
\end{tabular}
\caption{Comparison of the three MFs residuals for B98/B03 temperature
data (filled circles) to the analytical predictions with the best fit
value of $f_{\rm NL}$ for each functional (solid lines). The
analytical predictions are normalized to the maximum value of the
Gaussian part while the data points are normalized to the maximum of
the Monte Carlo average. The left figure is for HEALPix
$N_\mathrm{side}=256$ resolution while the right figure is for
$N_\mathrm{side}=512$. From left to right in each figure, we show the
$\ell_{\rm min}=$ 6, 12 and 20 cases. The error-bars represent the
standard deviation at 1-$\sigma$ estimated from 1000 Gaussian Monte
Carlo simulations.}
\label{fig3}
\end{figure*}

In Fig. \ref{fig3} we show the analytical non-Gaussian corrections
$\Delta V_J$ (eq.[\ref{eq:deltav}]) for each MF compared to the
residuals obtained by subtracting to the B98/B03 data MFs their Monte
Carlo average, that is, $V_J^{\rm B98+B03} - \bar{V}_J(f_{\rm
NL}=0)$. The error-bars represent the $1\sigma$ error estimated from
1000 Gaussian Monte Carlo simulations. The analytical residuals are
computed using the best fit value of $f_{\rm NL}$ as obtained by
minimizing the $\chi^2$ in the equation (\ref{chi2}), albeit this is
done separately for each MF, ignoring (only for the sake of this plot)
correlations among different functionals. The analytical $\Delta V_J$
in Fig.~\ref{fig3} are normalized to the maximum of their Gaussian
part, while the data points are normalized to the maximum of the Monte
Carlo average. We show results both for $N_\mathrm{side}=256$ (left)
and for $N_\mathrm{side}=512$ (right) and for a low-multiple cut at
$\ell_{\rm min}=$ 6, 12 and 20 (left to right). The agreement of the
residual plots suggests that it is safer to adopt the most
conservative cut at $\ell_{\rm min}=20$. Since increasing further the
cut does not appear to yield further advantage, we focus on a
conservative choice $\ell_{\rm min}=20$ for our final analysis.

Table \ref{table1} lists the confidence intervals for $f_{\rm NL}$
estimated from the BOOMERanG data at different $\ell_{\rm min}$ and
$\ell_{\rm max}$ thresholds. The results are obtained taking into
account the full covariance matrix of the $V_k$'s, as expressed in the
equation (\ref{chi2}) above. The MFs computed at different HEALPix
resolution encode different statistical information on the underlying
field and thereby combining two sets of MFs improve the limits on
$f_{\rm NL}$.  We build a global covariance matrix to take also into
account correlations among the two sets. We repeat this process for
each $\ell_{\rm min}$ value considered.

In the conservative case of $\ell_{\rm min}=20$ using the ``combined"
estimator, our $\chi^2$ analysis yields $-670<f_{\rm NL}<30$ at
1-$\sigma$ level and $-1020<f_{\rm NL}<390$ at 95\% CL, while the
minimum (best fit) value of $f_{\rm NL}$ is at $-320$. The previous
analysis performed only on the B03 dataset \citep{detroia} produced
limits weaker than those obtained in this paper by a factor $\simeq
1.4$. The improvement is due to a combination of both enhanced sky
coverage and the use of the analytical estimator.  The B98 dataset
alone has never been subject so far to $f_{\rm NL}$ analysis. We have
checked that this dataset too is compatible with $f_{\rm NL}=0$ at
1-$\sigma$ with $\Delta f_{\rm NL} = 470$ ($\Delta f_{\rm NL} = 930$
at 2-$\sigma$).

We can also quantify the cost of imposing a low $\ell$ cut to the
data. In fact, had we not considered an effective $\ell_{\rm min}$ value,
one would expect to reduce the confidence interval on $f_{\rm NL}$ by
$\simeq 1.6$ (c.f.\ the last row in Table~\ref{table1}, obviously
obtained not from the data but from a simulation containing a low
resolution pattern). In practice, this could be obtained by adding to
the dataset a low resolution CMB field coming, e.g., from the WMAP
data. While this would give us tighter constraints on $f_{\rm NL}$, we
prefer to focus here on the limits one can derive from the BOOMERanG
data alone. Note also, that a diminished sensitivity to low resolution
features is a characteristic common to most -if not all- of the
suborbital experiments. The accurate measurement of the CMB at low and
high multipoles with one single experiment is rather a prerogative of
space born missions, which enjoy the necessary stability and long term
integration capability. Our analysis is the first (to our knowledge)
to explicitly take into account this effect for a suborbital
experiment. To explain the significant broadening of $f_{\rm NL}$
constraints caused by the low $\ell$ cut, one can note that for the
underlying (``local'') form of non-Gaussianity we are probing here,
the low multipoles are actually very important. In fact, most of the
signal in the reduced bispectrum $b_{\ell_1\ell_2\ell_3}$ lies in
``squeezed'' $\ell$-space triangles, with one side much smaller than
the other two. When probing non-Gaussianity one is basically comparing
signal at the lowest multipole with two of the highest multipoles. As
a result, S/N increases as $\ell_{\rm max}/\ell_{\rm min}$ so one can either
increase $\ell_{\rm max}$ for a given $\ell_{\rm min}$ (which explains, e.g.,
the improvement of WMAP over COBE and the forecasted improvement of
\textsc{Planck} over WMAP), or reducing $\ell_{\rm min}$ for a given
$\ell_{\rm max}$.
\begin{table}\begin{center}
\begin{tabular}{|c|c|c|c}
\hline
\hline
 $\ell_{\rm min}$ & $\ell_{\rm max}=512$ & $\ell_{\rm max}=1000$  &  Combined \\
&                       $1\sigma$\hspace{0.5cm} $2\sigma$ & $1\sigma$\hspace{0.5cm} $2\sigma$ &$1\sigma$\hspace{0.5cm} $2\sigma$ \\
\hline
6 & 340 \hspace{0.5cm} 660 & 790 \hspace{0.5cm} 1570  & 320 \hspace{0.5cm} 620 \\
\hline
12 & 360 \hspace{0.5cm} 710 & 970 \hspace{0.5cm} 1930  & 350 \hspace{0.5cm} 690 \\
\hline
20 & 380 \hspace{0.5cm} 730 & 910 \hspace{0.5cm} 1830  & 360 \hspace{0.5cm} 730 \\
\hline
2  & 260 \hspace{0.5cm} 510 & 470 \hspace{0.5cm} 920  & 260 \hspace{0.5cm} 510 \\
\hline
\hline
\end{tabular}
\caption{The confidence intervals for $f_{\rm NL}$ estimated for the
B03/B98 data with different low and high $\ell$ cut values (see
text). We show both the $1\sigma$ and $2\sigma$ confidence
interval. The last two columns are obtained from a combined analysis
of data for the two $\ell_{\rm max}$ values. The last row is derived
using a simulation with no effective low $\ell$ cut and shows the
improvement that could be obtained if a low resolution pattern had
been present in the BOOMERanG field.}
\label{table1}
\end{center}
\end{table}

\section{Conclusions}
\label{4}

We have analysed data from the BOOMERanG experiment, combining for the
first time the temperature maps of the 1998 and 2003 campaigns, to
constrain a non-Gaussian primordial component in the observed CMB
field. We focused on Minkowski Functionals, comparing the data to
analytical perturbative corrections in order to get constraints on the
non-linear coupling parameter $f_{\rm NL}$. We have used a set of
highly realistic simulation maps of the observed field generated
assuming a Gaussian CMB sky, since the formalism we have adopted does
not require non-Gaussian simulation maps. We studied the effect that
the lack of low resolution CMB features in the BOOMERanG data has on
$f_{\rm NL}$ constraints. We find $-670<f_{\rm NL}<30$ at $68\%$ CL
and $-1020<f_{\rm NL}<390$ at 95\% CL. These limits are significantly
better than those published in a previous analysis limited to the
BOOOMERanG 2003 data ($-800<f_{\rm NL}<1050$ at 95\% CL), represent
the best results to date for suborbital experiments and probe angular
scales smaller than those accessible to the WMAP.

\section*{Acknowledgments}
The BOOMERanG team gratefully acknowledge support from the CIAR, CSA,
and NSERC in Canada; Agenzia Spaziale Italiana, 
University La Sapienza and Programma Nazionale Ricerche
in Antartide in Italy; PPARC and the Leverhulme Trust in the UK; and
NASA (awards NAG5-9251 and NAG5-12723) and NSF (awards OPP-9980654 and
OPP-0407592) in the USA. Additional support for detector development
was provided by CIT and JPL. Field, logistical, and flight support
were supplied by USAP and NSBF; This research used resources at NERSC,
supported by the DOE under Contract No. DE-AC03-76SF00098, and at
CASPUR (Rome, Italy: special thanks are due to M.\ Botti and F.\
Massaioli). We also acknowledge partial support from ASI Contract
I/016/07/0 ``COFIS'' and ASI Contract Planck LFI activity of Phase E2.
Some of the results in this paper have been derived using
the HEALPix package~\citep{healpix}. C.~H. acknowledges support from
the Particle Physics and Astronomy Research Council grant number
PP/C501692/1 and a JSPS (Japan Society for the Promotion of Science)
fellowship.


\end{document}